\documentclass[aip,rsi,amsmath,amssymb,reprint,prl]{revtex4-1}
\usepackage{graphicx}
\usepackage[T2A]{fontenc}
\usepackage{dcolumn}
\usepackage{bm}

\begin{document}
\preprint{AIP/123-QED}

\title[Limits of sub-Doppler cooling for atoms with various recoil parameter]{Limits of sub-Doppler cooling for atoms with various recoil parameter}


\author{A.A.Kirpichnikova}
\affiliation{
Institute of Laser Physics, 630090, Novosibirsk, Russia 
}%

\author{O.N.Prudnikov}
\email{oleg.nsu@gmail.com}
\author{R.Ya Il'enkov}
\author{A.V. Taichenachev}

\affiliation{
Institute of Laser Physics, 630090, Novosibirsk, Russia 
}%
\affiliation{
Novosibirsk State University, 630090, Novosibirsk, Russia 
}%

\author{V.I. Yudin}%
\affiliation{
Institute of Laser Physics, 630090, Novosibirsk, Russia 
}%
\affiliation{
Novosibirsk State University, 630090, Novosibirsk, Russia 
}%
\affiliation{Novosibirsk State Technical University, 630073,
Novosibirsk, Russia }

\date{\today}

\begin{abstract}
We perform detailed analysis of sub-Doppler cooling limits for
various atoms by direct solving quantum kinetic equation for atom
density matrix in standing-wave light field generated by
counterpropagating waves. It was demonstrated that the polarization
gradient cooling effects are sensitive to atom recoil parameter (the
ratio of recoil energy to natural linewidth) that results to
limitation of sub-Doppler cooling and allows to outline the limits
of well-known sub-Doppler cooling theory. We also give a comparison
the cooling limits for well-known $\sigma_+-\sigma_-$ and $lin\perp
lin$ configurations.
\end{abstract}



\maketitle

\section{Introduction}
Since the mid-1980s the laser cooling of atoms was rapidly
developing field of laser and atomic physics. Nowadays laser cooled
atoms are widely used in ultrahigh-resolution spectroscopy,
developing new generation of time and frequency standards
\cite{lundow,taich_ufn,marti}, achieving Bose-Einstein condensation
of neutral atoms \cite{cornell,ketterle}, simulating models of
quantum effects in condensed matter and interatomic collision
 investigation \cite{garraway,bloch}.

To this day various approaches to describe laser cooling have been
developed. At the initial stage of research the semiclassical
approaches were widely used
\cite{min,kaz,metcalf,dal1985,Javanainen1991,prudnikov1999,prudnikov2003,arimondo2004}.
These approaches allow describing the kinetics of atoms in terms of
diffusion and forces acting on atoms, resulted from recoil processes
due to absorption or emission of light field photons.

Semiclassical approaches are principally limited by a small value of
a momentum transmitted to atoms from a single light field photon in
comparison with the width of atom momentum distribution, $\hbar
k/\Delta p \ll 1$, as well as by a small value of recoil parameter
$\varepsilon_R = \omega_R/\gamma \ll 1$, that is the ratio of recoil
energy $\hbar \omega_R = \hbar^2 k^2/2M$ to the natural width of
optical transition $\gamma$ used for laser cooling. Smallness of
these two parameters ($\varepsilon_R$ and $\hbar k/\Delta p$) allows
separating a fast evolution of internal degrees of freedom from a
slow evolution of translational degrees of atoms. In this case the
complex quantum kinetic equation for atom density matrix can be
reduced to the Fokker-Plank equation for distribution function in
coordinate and momentum spaces with the force acting on atom and
diffusion coefficients
\cite{dal1985,Javanainen1991,prudnikov1999,prudnikov2003}. The
developing of semiclassical theory was of great importance to
understand the basic principles of the laser cooling including the
Doppler cooling \cite{min,kaz,metcalf,itano} and the sub-Doppler
cooling \cite{metcalf,dal1989,weiss1989,pru1999} mechanisms  in
optical molasses, i.e. in light fields formed by the pairs of
coutrerpropagating waves.

%

An alternative way is to use quantum approaches allowing taking into
account all recoil effects in interaction of atoms with resonance
light field photons. Theoretical describing of such processes is
quite complicated because the interaction between atoms and photons
brings changing of both internal and motional degrees of freedom.
In particular, instead of solving quantum kinetic equation for atom density matrix
statistical Monte Carlo wave-function method was developed
\cite{molmer92,Castin1996}. The number of variables involved in
wave-function simulation is determined by the relevant Hilbert space
dimension N much smaller than the one required for calculations with
density matrices ($\sim N^2$). However, it spends much time for
computer to simulate and obtain appropriate statistics for
calculating mean over the atom trajectories.

Kinetic quantum equation for atom density matrix contains much more
variables than wave-function approach and at the beginning of
investigation various approximation was used to solve it. For
example, to describe limits of sub-Doppler laser cooling in $lin
\perp lin$ field \cite{dal1991,dal1991m} approximation of low
intensity limit was used allowing applying simplified equation for
ground state atom density matrix. Moreover, in the papers mentioned above, it was used
the secular approximation,  $\sqrt{U_0/\hbar \omega_R}\ll
|\delta|/\gamma$, supposing a gap between energy bands of optical
potential to be much wider than bandwidth. Light shift $U_0$ is
determined by the depth of optical potential, $\delta=
\omega-\omega_0$ is detuning parameter of light field frequency
$\omega$ from atomic resonance frequency $\omega_0$. In case of
absence of optical potential, i.e. in a light field formed by a pair of
counterpropogating waves with orthogonal circular polarizations
$\sigma_+-\sigma_-$, approaches developed in the papers
\cite{dal1991,dal1991m} can not be used. However, the methods of
p-families could be used in such cases \cite{borde,aspect1989,
Castin1989,pru2003}.

In the papers \cite{pru2007,pru2007jetp,pru2011}, we offered an
universal quantum approach allowing describing steady-state of atoms in
1D configuration formed by
counterpropogating waves with arbitrary elliptical
polarization interacting with arbitrary optical transition $Jg \to Je$.
This technique appears to be more general and face no
restriction mentioned above. Moreover, in case of insufficient
smallness of recoil parameter $\varepsilon_R$, the recoil effects
were noted to become more sufficient, and significant discrepancy in
temperatures was observed in contrast with predictions made by
semiclassical and quantum theories
\cite{Castin1989,Chalony2011,pru2019,pru2015,kalg}.

In this paper, we are performing detailed analysis of the
limits of polarization-gradient sub-Doppler cooling of atoms with
different ratio of recoil energy to the natural linewidth varied in
wide range from extremely small values when well-known semiclassical
and quantum models work properly to values close to $\varepsilon_R
\simeq 1$ when the recoil effects carry critical weight.
Our analysis allows to underline the limits of well-known
sub-Doppler cooling theories for different polarization-gradient
field configurations $\sigma_+-\sigma_-$ and $lin\perp lin$.

\section{Master equation}
Let us consider the laser cooling of atoms  with closed optical
transition $Jg \to Je$ where $Jg$ and $Je$ are angular momenta
of the ground ($g$) and excited ($e$) states. The atoms are
resonantly interacting with 1D field formed by
counterpropagating light waves along z axis:
\begin{equation}
\label{lightfield}
 {\bf E}(z,t) = E_0 \left({\bf e}_1e^{ikz}+{\bf
e}_2e^{-ikz}\right)e^{-i\omega t}+c.c.
\end{equation}
Here  $E_0$ is the complex amplitude of the light waves, $\omega$ is
the field frequency and $k = \omega/c$ is the wavevector. The
polarization vectors ${\bf e}_1$ and ${\bf e}_2$ could
be expressed in complex circular basis ${\bf e}_{\pm}  = \mp ({\bf
e}_x \pm {\bf e}_y)/\sqrt{2}$ and ${\bf e}_0 = {\bf e}_z$
\begin{equation}
{\bf e}_n = \sum_{\sigma = 0,\pm 1} e^{\sigma}_{n}{\bf e}_{\sigma}
\, , \,\,\, n = 1,2 \,.
\end{equation}
Here we will consider the following two configurations:
\begin{itemize}
    \item $lin\perp lin$  configuration is defined by ${\bf e}_1$ and ${\bf
    e}_2$ with orthogonal linear polarization \cite{dal1989}.
    \item $\sigma_+-\sigma_-$  configuration is defined by two orthogonal circular polarization of ${\bf e}_1  = {\bf e}_+$ and ${\bf
    e}_2= {\bf e}_-$ \cite{dal1989}.
\end{itemize}
These simple configurations represent the light field with only one
parameter having spatial dependence: the local ellipticity for $lin
\perp lin$ configuration and the local polarization orientation
angle for $\sigma_+-\sigma_-$ configuration. The other field
parameters like field intensity and phase have no spatial dependence
fore these two configurations. Note that the general 1D field
configuration with several parameters varied along the propagation
axes can be represented by $\varepsilon_1 - \theta - \varepsilon_2$
field configuration formed by counterpropagating plane waves with
arbitrary elliptical polarizations
\cite{prudnikov1999,pru1999,pru2001,pru2016}.

The kinetic evolution of a low-density atomic ensemble (when
interatomic interaction could be neglected) is described by the
quantum kinetic equation for the atomic density matrix ${\hat \rho}$
in the single-particle approximation
\begin{equation} \label{kin}
\frac{\partial {\hat \rho}}{\partial t} = -\frac{i}{\hbar} \left[
{\hat H}, {\hat \rho}\right] +{\hat \Gamma}\{{\hat \rho}\} \,,
\end{equation}
where ${\hat H}$ is the Hamiltonian, and the term ${\hat
\Gamma}\{{\hat \rho}\}$ describes the relaxation in the process of
spontaneous decay. The Hamiltonian can be divided into the sum
\begin{equation}\label{hamiltonian}
{\hat H} = \frac{{\hat p}^2}{2M} +{\hat H}_0 +{\hat V} \, ,
\end{equation}
where the first term is the kinetic energy operator; ${\hat H}_0 =
-\hbar \delta {\hat P}_e$ is the Hamiltonian of the free atom in
rotating wave approximation (RWA); $\delta = \omega-\omega_0$ is
light field detuning from the atomic transition frequency
$\omega_0$; and
\begin{equation}
{\hat P}_e = \sum_{\mu} |J_e,\mu \rangle \langle J_e, \mu |
\end{equation}
is the projection operator to the exited state sublevels $|J_e,\mu
\rangle$ where  $\mu$ is the angular momentum projection on the quantization axis ( $-Je\leq\mu \leq Je$ ). The last
term in (\ref{hamiltonian}) ${\hat V}$ describes the atom-light interaction
which in electric dipole approximation takes a form
\begin{eqnarray}
{\hat V} &= & {\hat V}_1 \exp(ikz) +{\hat V}_2 \exp(-ikz) \nonumber \\
{\hat V}_n &= & \frac{\hbar \Omega}{2} \left({\hat {\bf D}} \cdot
{\bf e}_n \right) =  \frac{\hbar \Omega}{2} \sum_{\sigma = 0, \pm 1}
{\hat D}_{\sigma} e^{\sigma}_{n}\, \,\,\,\, n = 1,2\,.
\end{eqnarray}
Here $\Omega$ is the Rabi frequency. The circular components of
operator ${\hat D}$ are expressed via the Clebsch-Gordan
coefficients according to Wigner-Eckart theorem:
\begin{equation}
{\hat D}_{\sigma} = \sum_{\mu,m} C^{Je,\mu}_{Jg,m ;\,
1,\sigma}\,|J_e,\mu \rangle \langle J_g,m|\,.
\end{equation}
The last term of the kinetic equation (\ref{kin}) describing the
relaxation due to spontaneous decay
taking into account photon recoil has a well-known form (see for
example \cite{pru2011}):
\begin{eqnarray}
\lefteqn{{\hat \Gamma}\{{\hat \rho}\}=-\frac{\gamma}{2}\left({\hat P}_e
{\hat \rho}+{\hat \rho}{\hat P}_e \right)}\nonumber \\
&\mbox{}+ \frac{3}{2}\gamma\,\left \langle \sum_{\xi = 1,2}
\left({\hat {\bf D}}\cdot {\bf e}_{\xi}({\bf k}) \right)^{\dagger}
e^{-i{\bf k}{\hat {\bf r}}}{\hat \rho}\, e^{i{\bf k}{\hat {\bf r}}}
\left({\hat {\bf D}}\cdot {\bf e}_{\xi}({\bf k}) \right) \right
\rangle_{\Omega_{\bf k}}\, ,
\end{eqnarray}
where $\langle \ldots \rangle_{\Omega_{\bf k}}$ denotes averaging
over the directions of emission of spontaneous photons having a
momentum $\hbar {\bf k}$ with two orthogonal polarizations ${\bf
e}_{\xi}({\bf k})$.

To solve the kinetics of laser cooling it is convenient to
use the coordinate representation for density matrix in which the
spontaneous relaxation operator accounting recoil effects in 1D
geometry takes the simplest form:
\begin{eqnarray}
{\hat \Gamma}\left\{{\hat \rho}(z_1,z_2)\right\}&=&
-\frac{\gamma}{2}\left({\hat P}_e{\hat \rho}(z_1,z_2)+{\hat
\rho}(z_1,z_2){\hat P}_e\right)  \nonumber \\
& &+\gamma \sum_{\sigma = 0,\pm1}\kappa_{\sigma}(q) {\hat
D}^{\dagger}_{\sigma} {\hat \rho}(z_1,z_2) {\hat D}_{\sigma}\,
\end{eqnarray}
where $q = z_1-z_2$ and functions $\kappa_{0,\pm 1}$ are
\begin{eqnarray}
\kappa_0(q) &=& 3\left(\frac{\sin(kq)}{(kq)^3}-\frac{\cos(kq)}{(kq)^2} \right) \, \nonumber \\
\kappa_{\pm1}(q) &=&
\frac{3}{2}\left(\frac{\cos(kq)}{(kq)^2}+\frac{\sin(kq)}{kq}-\frac{\sin(kq)}{(kq)^3}
\right) \, .
\end{eqnarray}

For solving equation (\ref{kin}) in steady-state and analysis of the
limits of sub-Doppler cooling we are using a generalized continious
fraction method suggested by us and described in detail using Wigner
in \cite{pru2007,pru2007jetp} and coordinate representations in
\cite{pru2011} for atomic density matrix. The steady-state solution of
(\ref{kin}) occurs to be periodic in coordinate $z$ and allows
factorizing on spacial harmonics:
\begin{equation}
{\hat \rho}(z,q) = \sum_n {\hat \rho}^{(n)}(q)\,e^{inkz}.
\end{equation}
Thus the task is reduced to calculating of the amplitudes ${\hat \rho}^{(n)}(q)$.
The equation for steady-sate Fourier harmonics ${\hat \rho}^{(n)}$
can be written in recurrent form with three terms:
\begin{equation}
-n \frac{i}{M} \frac{\partial}{\partial q}{\hat \rho}^{(n)} =
{\mathcal L}_0 \lbrace {\hat \rho}^{(n)} \rbrace + {\mathcal L}_+
\lbrace {\hat \rho}^{(n-1)} \rbrace + {\mathcal L}_- \lbrace {\hat
\rho}^{(n+1)} \rbrace \,,
\end{equation}
where operators ${\mathcal L}$ are
\begin{eqnarray}
{\mathcal L}_+ \{ {\hat \rho} \} &=& -\frac{i}{\hbar} \left( {\hat
W}_1 {\hat \rho} e^{ikq/2}
- {\hat \rho}{\hat W}_1 e^{-ikq/2} \right) \, \nonumber \\
{\mathcal L}_- \{ {\hat \rho} \} &=& -\frac{i}{\hbar} \left( {\hat
W}_2 {\hat \rho} e^{-ikq/2}
- {\hat \rho}{\hat W}_2 e^{ikq/2} \right) \, \nonumber \\
{\mathcal L}_0 \{ {\hat \rho} \} &=& -\frac{i}{\hbar} \left( {\hat
H}_0 {\hat \rho} - {\hat \rho}{\hat H}_0\right) - {\hat \Gamma} \{
{\hat \rho} \}
\end{eqnarray}
with matrix coefficients
\begin{equation}
{\hat W}_1 = \left(
\begin{array}{cc}
    0 & {\hat V}_1 \\ {\hat V}^{\dagger}_2 & 0
\end{array} \right) , \,
{\hat W}_2 = \left(
\begin{array}{cc}
    0 & {\hat V}_2 \\ {\hat V}^{\dagger}_1 & 0
\end{array} \right) .
\end{equation}

Note that harmonics of atomic density matrix $\hat{\rho}^{(n)}$
depend on variable $q$ and contain the information on quantum
correlations of atomic states between two points separated in space
$z_1 = z +q/2$ and $z_2 = z - q/2$. As far as the correlation should
decay with growing $|q|$, we can cut it with large enough value $q_{max}$. In
a Wigner representation $q_{max}$ defines detailization in momentum
space $\Delta p \simeq \pi/q_{max}$. In our simulations, we typically
used $q_{max} \leq 10/k$, however, for some parameters of laser
cooling the quantum correlation length is found to be quite large
and requiring increasing $q_{max}$ up to $ \sim 100/k$ to account
for these effects correctly.

\section{Steady-state of laser cooling}
The steady-state solution of quantum kinetic equation (\ref{kin})
is determined by the light field and atomic parameters. The light
field parameters are: the intensity of light waves $I$, optical
frequency $\omega$, and the spatial polarization configuration which is defined, as was noted above,
by the polarizations of opposite light
waves. The atomic parameters are the type of optical transition (which is defined by the
total angular momenta of the ground and the exited states $J_g \to
J_e$), the transition frequency $\omega_0$,
the dipole moment of optical transition $d$, the natural linewidth
$\gamma$, and the atom mass $M$.

So, within the above list of parameters a few
dimensionless parameters determined the steady-state solutions of
master equation (\ref{kin}) can be separated. These are
\begin{itemize}
    \item $\delta/\gamma$ is dimensionless detuning
    \item $\Omega/\gamma$ is dimensionless Rabi frequency containing
    information on the light waves intensity $\Omega/\gamma =
    \sqrt{I/2I_{sat}}$ ($I_{sat}=2\pi^2 \gamma\hbar c/\lambda^3$ is determined by the dipole momentum of optical transition, see for example \cite{adams})
    \item $\varepsilon_R = \omega_R/\gamma$ is recoil
    parameter
\end{itemize}
As well, the light field polarization configuration and the type of
optical transition $J_g \to J_e$ stay as additional parameters into
represented list of dimensionless parameters of laser cooling task.

\begin{table}[tbp]
\begin{center}
\caption{Recoil and optical transitions parameters used for laser
cooling of different atoms. }\label{optical_transitions}
\begin{tabular}{|c|c|c|c|c|}
\hline
atom  &cooling optical&$\gamma/2\pi$&$\lambda$& $\varepsilon_R$\\
            & transition &          (MHz) &            (nm) &             \\ \hline
$^{174}$Yb &$6^1S_{0} \to 6^1P_{1}$   &28  &399  &$3\cdot 10^{-4}$ \\
                     &$6^1S_{0} \to 6^3P_{1}$   &0.18  &556  &$2\cdot 10^{-2}$\\ \hline
$^{87}$Sr    &$5^1S_{0} \to 5^1P_{1}$   &32  &461  &$3\cdot 10^{-4}$ \\
                     &$5^1S_{0} \to 5^3P_{1}$   &$0.7\cdot 10^{-3}$  &689  &0.6\\ \hline
$^{40}$Ca   &$4^1S_{0} \to 4^1P_{1}$   &34.2  &422.8  &$0.8\cdot
10^{-3}$\\
&$3^1S_{0} \to 3^3P_{1}$   &$4\cdot 10^{-4}$  &657.3  &$28$
\\ \hline
$^{24}$Mg   &$3^1S_{0} \to 3^1P_{1}$   &78 &  285.3  &$1.3\cdot 10^{-3}$ \\
&$3^3P_{2} \to 3^3D_{3}$   &26.7  &383.9  &$2.1\cdot10^{-3}$ \\
&$3^1S_{0} \to 3^3P_{1}$   &$31.2\cdot 10^{-6}$  &457  &$1.2\cdot
10^3$ \\ \hline $^{133}$Cs   & $6^2S_{1/2} \to 6^2P_{3/2}$   &5
&852.35  &$4\cdot 10^{-4}$\\ \hline $^{85}$Rb    &$5^2S_{1/2} \to
5^2P_{3/2}$   &5.9  &780.24  &$6\cdot 10^{-4}$\\ \hline $^{39}$K
&$4^2S_{1/2} \to 4^2P_{3/2}$   &6.2   &766.7   &$1.4\cdot 10^{-3}$\\
\hline $^{23}$Na    &$3^2S_{1/2} \to 3^2P_{3/2}$   &9.9  &589.16
&$2.5\cdot 10^{-3}$\\ \hline $^{7}$L         &$2^2S_{1/2} \to
2^2P_{3/2}$   &5.9  &670.96   &$10^{-2}$\\ \hline $^{1}$H
&$1^2S_{1/2} \to 2^2P_{3/2}$   &99.58   &121.57   &0.13 \\ \hline
$^{27}$Al      &$3^2P_{3/2} \to 3^2D_{5/2}$  &13    & 309.4& $6\cdot
10^{-3}$\\ \hline $^{52}$Cr      & $a^7S_{3} \to z^7P_{4}$ &5 &425.6
&$4\cdot 10^{-3}$ \\ \hline $^{56}$Fe &$a\,^5D_{4} \to z\, ^5P_{5}$
&2.58   &372  &$10^{-2}$\\ \hline $^{69,71}$G
&$4^2P_{3/2} \to 4^2D_{5/2}$   &25  & 294.4 &  $1.3\cdot 10^{-3}$\\
\hline $^{107}$Ag  &$5^2S_{1/2} \to 5^2P_{3/2}$   &2.2  & 328 &
$0.8\cdot 10^{-3}$\\ \hline $^{115}$In   &$5^2P_{3/2} \to
5^2D_{5/2}$   &20.7  &325.7   &$0.8\cdot 10^{-3}$\\ \hline
$^{199}$Hg  &$6^1S_{0} \to 6^3P_{1}$   &1.32  &253.7   &$1.3\cdot 10^{-2}$ \\
\hline
\end{tabular}
\end{center}
\end{table}

The third parameter $\varepsilon_R = \hbar k^2/(2M\gamma)$  is
determined by atomic mass and optical transition used for
realization of laser cooling (see the table
\ref{optical_transitions}). It is small enough $\varepsilon_R\ll 1$
and varies form $\sim 10^{-4} - 10^{-1}$ for majority of atomic
optical transitions used for laser cooling. It tends to reach
extremely low values $\varepsilon_R< 10^{-3}$ for alkaline elements
(like Cs and Rb are cooling with the use of D2 line). As well, it
gets the values above $10^{-1}$ for narrow-line optical transition,
for example, intercombination transitions $^1S_0-^3P_1$ of Sr, Ca,
and Mg atoms.

Bellow we are demonstrating the influence of recoil parameter on the
results of laser cooling in the case of $\varepsilon_R$ being not
extremely small and outlining the limits of sub-Doppler cooling for
intermediate values of the recoil parameter ($ 10^{-3}\leq \varepsilon_R \leq 10^{-1} $).

Emphasis should be placed on fact that in a field of low intensity at secular approximation
of papers \cite{dal1991,dal1991m}, describing Sisyphus cooling in a
field of $lin \perp lin$ configuration, the only one parameter
\begin{equation}
U_0 =
\frac{|\delta|}{3\,\omega_R}\frac{|\Omega|^2}{\left(\delta^{2}+\gamma^{2}/4\right)}
\end{equation}
 defines the steady state solution. Therefore, in our analysis, we are using
 this parameter  instead of $\Omega/\gamma$. This will allow us to compare our results with the earlier results
 in the papers
\cite{dal1991,dal1991m} obtained with some approximations. Thus the list of dimentionless parameters is following: $\delta/\gamma$, $U_0$, $\varepsilon_R$.

\subsection{Extremely low recoil parameter $\varepsilon_R\leq 10^{-3}$}
 In this section we demonstrate the results of steady-state
solution of master equation (\ref{kin}) for extremely small recoil
parameter $\varepsilon_R\leq 10^{-3}$. This limit is well studded by
many authors with the use of semiclassical approaches
\cite{min,kaz,metcalf,dal1985,Javanainen1991,prudnikov1999,prudnikov2003,arimondo2004}
as well as the quantum approaches \cite{dal1991,dal1991m}. We made
our analysis for transition $J_g=1\to J_e=2$ admitting the
sub-Doppler laser cooling for both types of light field
configurations $lin\perp lin$ and $\sigma_+-\sigma_-$
\cite{dal1989}. It is the simplest example allowing us to compare
the cooling limits for these configurations.

\begin{figure}[h]
\centerline{\includegraphics[width=3.2 in]{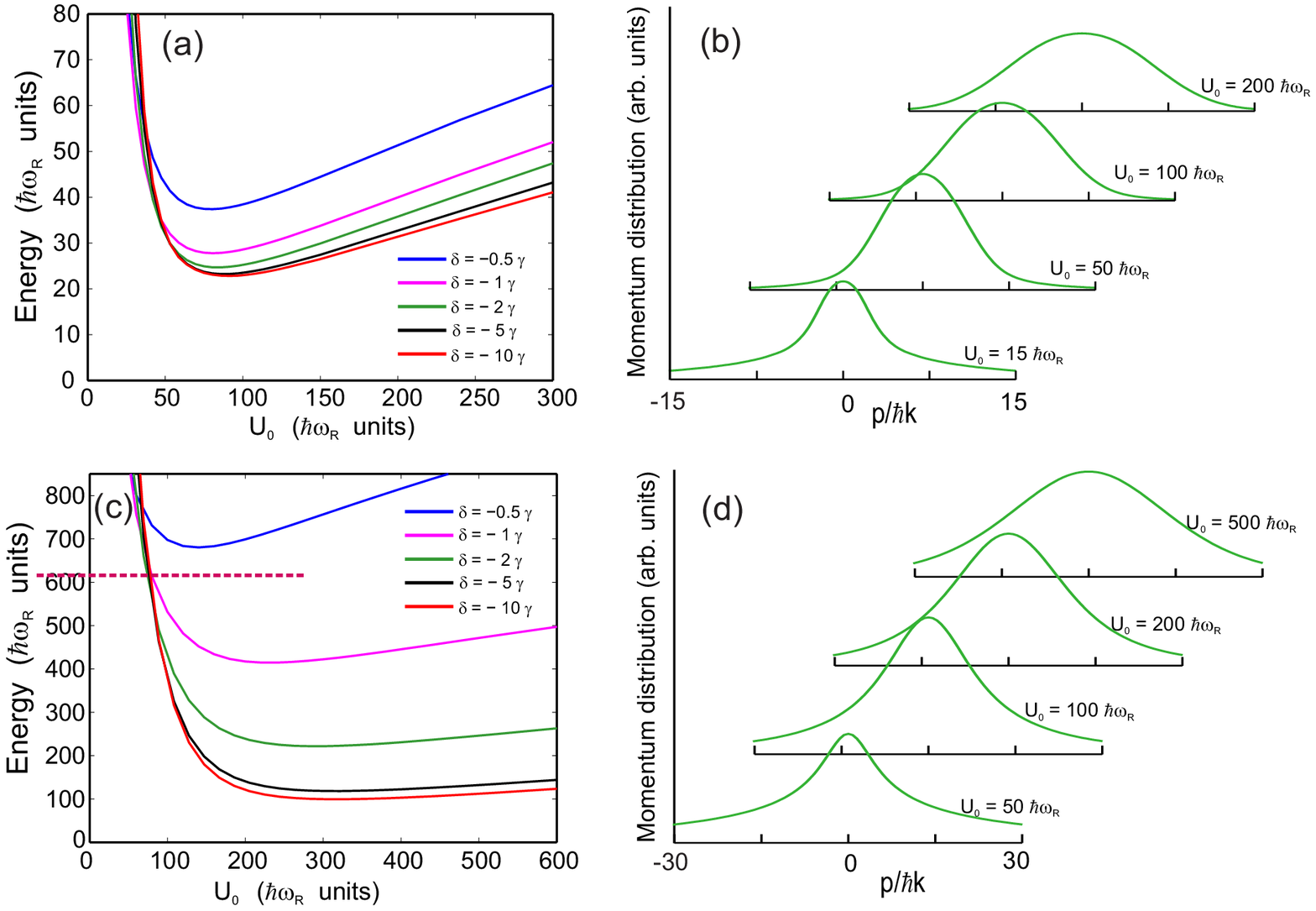}} \caption{(a)
Average kinetic energy of cold atoms as function of $U_0$ for
different detuning~$\delta$ in $lin\perp lin$ field (a) and in
$\sigma_+-\sigma_-$ field (c). Momentum distribution in of cold
atoms for different~$U_0$ at $\delta=-2\gamma$ in $lin\perp lin$
field (a) and $\sigma_+-\sigma_-$ field (d). Recoil parameter
$\varepsilon_R = 4\times10^{-4}$. Red dashed horizontal lines
correspond to average kinetic energy of atoms with Gaussian momentum
distribution and Doppler temperature $k_B T_D = \hbar \gamma/2$.}
\label{fig1}
\end{figure}

\begin{figure}[h]
\centerline{\includegraphics[width=3.2 in]{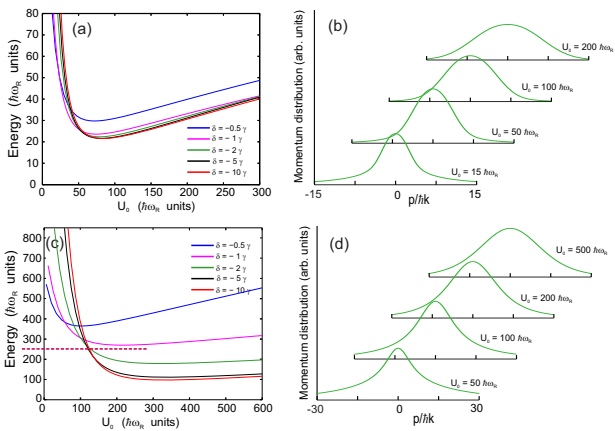}} \caption{(a)
Average kinetic energy of cold atoms as function of $U_0$ for
different detuning~$\delta$ in $\sigma_=-\sigma_-$ field (a) and in
$\sigma_+-\sigma_-$ field (c). Momentum distribution in of cold
atoms for different~$U_0$ at $\delta=-2\gamma$ in $lin\perp lin$
field (a) and $\sigma_+-\sigma_-$ field (d). Recoil parameter
$\varepsilon_R = 10^{-3}$. Red dashed horizontal lines correspond to
average kinetic energy of atoms with Gaussian momentum distribution
and Doppler temperature $k_B T_D = \hbar \gamma/2$.} \label{fig2}
\end{figure}

In the case of $lin\perp lin$ configuration, the result of average
kinetic energy of cooled atoms in steady-state are represented in
Fig.\ref{fig1}(a) for recoil parameters $\varepsilon_R =
4\times10^{-4}$ corresponding to Cs atoms cooling with use of D2
line, and for $\varepsilon_R = 10^{-3}$ in Fig.\ref{fig2}(a) for
comparison. Note that the average kinetic energy tends to have a
dependance only on $U_0$ parameter for enough large detunings that
corresponds to secular approximation, $\sqrt{U_0/\hbar \omega_R}\ll
|\delta|/\gamma$, used in \cite{dal1991,dal1991m}. The minimum
kinetic energy reaches the value $E_{min}\simeq 22\, \hbar \omega_R$
for $\varepsilon_R = 4\times10^{-4}$ and $E_{min}\simeq 20\, \hbar
\omega_R$ for $\varepsilon_R = 10^{-3}$ that is slightly less in
comparison with results of \cite{dal1991,dal1991m}. This discrepancy
arises from differences of optical transitions $J_g\to J_e$ under
consideration: $J_g=1\to J_e = 2$ in the our case, and $J_g = 1/2
\to J_e = 3/2$ in \cite{dal1991,dal1991m}. For the optical
transition $J_g = 1/2 \to J_e = 3/2$ our results are fully
consistent with the results of \cite{dal1991,dal1991m} for large
enough detunings \cite{pru2011}.

 The momentum distribution of cold atoms Fig.\ref{fig1}(b) and Fig.\ref{fig2}(b)
 occurs not to be Gaussian
functions and cannot be analyzed in terms of temperature, but in
terms of energy only. Indeed, the averaged kinetic energy,
$E_{kin}/\hbar \omega_R \simeq 1/(4\varepsilon_R)$, corresponding to
 equilibrium momentum distribution described by
Gaussian function with Doppler temperature $k_BT_D\simeq \hbar
\gamma/2$, is $E_{kin}/\hbar \omega_R \simeq 625$ for
Fig.\ref{fig1}(a) and $E_{kin}/\hbar \omega_R \simeq 250$ for
Fig.\ref{fig2}(a). One can see that it is much above the minimum
energy values in Fig.\ref{fig1}(a) and Fig.\ref{fig2}(a), that
corresponds to well known sub-Doppler cooling effects \cite{dal1989}
having semiclassical interpretation
\cite{metcalf,dal1985,Javanainen1991,prudnikov1999,prudnikov2003}.

We got the similar results for the steady-state in
$\sigma_+-\sigma_-$ light field configuration Fig.\ref{fig1}(c) and
Fig.\ref{fig2}(c). We have found that the average kinetic energy of
cold atoms also tends to have an unique dependence on $U_0$ for
large detunings (as for $lin \perp lin$). In these figures, the red
dashed horizontal lines correspond to the average kinetic energy of
atoms with Gaussian momentum distribution and Doppler temperature
$k_B T_D = \hbar \gamma/2$. As it seen, for atoms with recoil
parameter $\varepsilon_R$ smaller than $10^{-3}$, the universal
dependence on $U_0$ is achieved with enough large red detunings,
$|\delta|/\gamma \gtrsim 5$. Also, for these detunings the
sub-Doppler cooling gets minimum values.

\subsection{Not enough small value of recoil parameter}
\begin{figure}[h]
\centerline{\includegraphics[width=3.2 in]{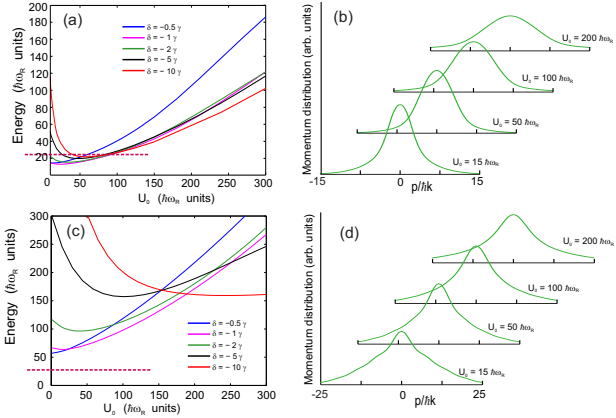}} \caption{(a)
Average kinetic energy of cold atoms as function of $U_0$ for
different detuning~$\delta$ in $lin\perp lin$ field (a) and in
$\sigma_+-\sigma_-$ field (c). Momentum distribution in of cold
atoms for different~$U_0$ at $\delta=-2\gamma$ in $lin\perp lin$
field (a) and $\sigma_+-\sigma_-$ field (d). Recoil parameter
$\varepsilon_R = 10^{-2}$. Red dashed horizontal lines correspond to
average kinetic energy of atoms with Gaussian momentum distribution
and Doppler temperature $k_B T_D = \hbar \gamma/2$.} \label{fig3}
\end{figure}

The more significant discrepancy from well-known picture of
sub-Doppler cooling theory we have found for the recoil parameters
at larger values ($\varepsilon_R > 10^{-3}$). For example, the
results of average kinetic energy as function of $U_0$ for different
detunings for $\varepsilon_R = 10^{-2}$ is shown in Fig.\ref{fig3}.
The red dashed horizontal lines correspond to average kinetic energy
of atoms with Gaussian momentum distribution and Doppler temperature
$k_B T_D = \hbar \gamma/2$. There is some area of $U_0$ and $\delta$
when the sub-Doppler cooling stays effective with use of light field
of $lin \perp lin$ configuration. However, the magneto-optical trap
does not operate in $lin\perp lin$ configuration.

For $\sigma_+-\sigma_-$ field, the sub-Doppler cooling mechanisms
lose their efficiency, when the average kinetic energy of cooled
atoms always stays above the Doppler cooling limit despite the fact
that all requirements for sub-Doppler cooling according to
well-known semiclassical theories
\cite{metcalf,dal1985,Javanainen1991,prudnikov1999,prudnikov2003}
are fulfilled, as well as the recoil parameter is small enough for
these theories to be valid, $\varepsilon_R \ll 1$.  As seen from
Fig.\ref{fig3}(c),
 the
$\sigma_+-\sigma_-$ configuration does not provide the sub-Doppler
cooling for these atoms. We suppose that our analysis explains the
experiments on Mg atoms cooling with use of $3s3p\, ^3P_2\to 3s3d\,
^3D_3$ optical transition \cite{Riedmann2012} and recent results of
Sr atoms cooling with use of $5s5p\, ^3P_2\to 5s4d\, ^3D_3$ optical
transition \cite{Hobson2019}, where temperatures above the Doppler
limit were achieved only.

\subsection{Narrow-line cooling}
\begin{figure}[h]
\centerline{\includegraphics[width=3.2 in]{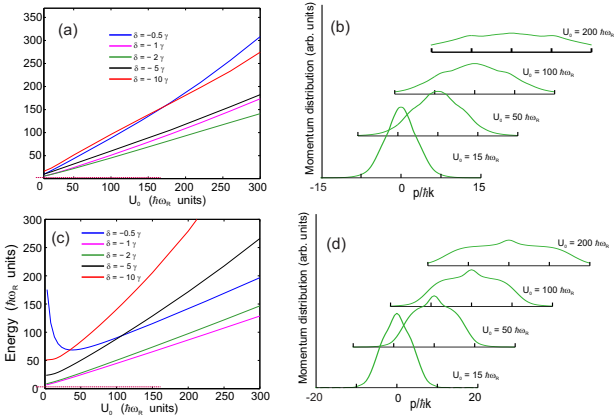}} \caption{(a)
Average kinetic energy of cold atoms as function of $U_0$ for
different detuning~$\delta$ in $lin\perp lin$ field (a) and in
$\sigma_+-\sigma_-$ field (c). Momentum distribution in of cold
atoms for different~$U_0$ at $\delta=-2\gamma$ in $lin\perp lin$
field (a) and $\sigma_+-\sigma_-$ field (d). Recoil parameter
$\varepsilon_R = 10^{-1}$. Red dashed horizontal lines correspond to
average kinetic energy of atoms with Gaussian momentum distribution
and Doppler temperature $k_B T_D = \hbar \gamma/2$.} \label{fig4}
\end{figure}

As it seen in Fig.\ref{fig4} for more larger values of recoil
parameter ($\varepsilon_R \simeq 10^{-1}$), the sub-Doppler cooling
effects becomes not effective even in $lin \perp lin$ field
configuration. In contrast to the case of extremely small values of
$\varepsilon_R$, the minimal average kinetic energy is achieved for
the smaller detunings $\delta \simeq -2\gamma$ in $lin\perp{lin}$
field and for $\delta \simeq -\gamma$ in $\sigma_+-\sigma_-$ field
for various values $U_0$.

\section{Conclusion}
We discussed the limits of sub-Doppler cooling of atoms in a fields
formed by couterpropagating waves with orthogonal linear or circular
polarizations. In our analysis we solved the master equation for
atom density matrix taking into account the quantum recoil effects
as well as the effects of saturation in light field. Our method
allows getting the steady-state solution without limitations and
approximations was used by other authors.

In our analysis we payed main attention to the cooling limits for
atoms with different ratio of recoil parameter $\varepsilon_R =
\omega_R/\gamma$. In particular, we show that the well-known picture
of sub-Doppler cooling is valid only in the limit of extremely small
values of $\varepsilon_R \leq 10^{-3}$. In this case, the average
kinetic energy of cooled atoms for large detunings tends to
dependence on light shift parameter $U$ only (for  $lin \perp lin$
as well as $\sigma_+-\sigma_-$ field configurations). For larger
values of $\varepsilon_R \simeq 10^{-2} - 10^{-1}$ the sub-Doppler
cooling mechanisms becomes less effective especially in
$\sigma_+-\sigma_-$ configuration usually used for magneto-optical
trap. As well, the minimum of kinetic energy of cooled atoms is
achieved for smaller red detunings $|\delta/\gamma|\simeq 1 $ in
comparison with $\varepsilon_R \leq 10^{-3}$ case ($|\delta/\gamma|
\gtrsim 5 $), i.e. in the case of $\varepsilon_R \geq 10^{-2}$ the
optimal detuning  becomes close to the optimal value $\delta/\gamma
= -1/2$ for Doppler cooling limit of two-level atom
\cite{min,kaz,metcalf}.

In addition, we emphasize that the steady-state of cooled atoms in
general is essentially non-equilibrium \cite{prujetpL} and therefore
unable to be described in term of unit temperature. In  low
intensity regime of laser cooling, the ``hot'' and ``cold''
fractions of atoms can be described by different temperatures of
these fractions \cite{kalg}. In this case, we can see a small
portion of atoms with sub-Doppler temperatures while the major atoms
is in ``hot'' fraction with temperature near the Doppler limit (see
Figs.\ref{fig1} - \ref{fig4}). Thus, for atoms with recoil parameter
above $\varepsilon_R \geq 10^{-2}$ the fraction of ``cold'' atoms
becomes nonessential and the main role in laser cooling plays the
well-known Doppler laser cooling mechanism, despite of the Zeeman
atomic degeneracy and presence of polarization gradient of the
cooling field.

\begin{acknowledgments}
The work is supported by RFBR 19-29-11014. The work of O.N.
Prudnikov is supported by RFBR and Novosibirsk region government
within the project 18-42-540003.  The work of R.Ya Il'enkov with
simulation for extremely low recoil parameter was supported by RFBR
19-02-00514. V.I Yudin was supported by RFBR 20-02-00505A.
\end{acknowledgments}


\end{document}